\documentclass[letterpaper, 10pt, conference]{IEEEtran}  
\usepackage{times}

\makeatletter
\let\NAT@parse\undefined
\makeatother

\usepackage[sort&compress,numbers,merge]{natbib}

\usepackage{multicol}

\usepackage{pgf} 

\usepackage{amsmath,subcaption}
\usepackage{theorem}
\usepackage{latexsym,amssymb,amsmath,amsfonts}
\usepackage{verbatim}
\usepackage{algorithm}

\usepackage{graphicx}
\usepackage{float}
\usepackage{flafter}
\usepackage[english]{babel}
\usepackage[utf8]{inputenc}
\usepackage{algpseudocode}

\renewcommand{\eqref}[1]{(\ref{eq:#1})}

\theoremstyle{plain} \theorembodyfont{\upshape}

\usepackage[normalem]{ulem} 

\IEEEoverridecommandlockouts                              
\title{\bf
Intent Communication between Autonomous Vehicles and Pedestrians}

\author{Milecia Matthews, ~Girish V. Chowdhary, Emily Kieson
  \thanks{Milecia Matthews, and  Emily Kieson are with
    Mechanical and Aerospace Eningeering at Oklahoma State University, Stillwater OK, Girish Chowdhary is with University of Illinois at Urbana Champaign from July 2016, {girishc@illinois.edu}. 
    }}

\begin{document}
\maketitle

\begin{abstract}
When pedestrians encounter vehicles, they typically stop and wait for a signal from the driver to either cross or wait. What happens when the car is autonomous and there isn't a human driver to signal them? This paper seeks to address this issue with an intent communication system (ICS) that acts in place of a human driver. This intent system has been developed to take into account the psychology behind what pedestrians are familiar with and what they expect from machines. The system integrates those expectations into the design of physical systems and mathematical algorithms. The goal of the system is to ensure that communication is simple, yet effective without leaving pedestrians with a sense of distrust in autonomous vehicles. To validate the ICS, two types of experiments have been run: field tests with an autonomous vehicle to determine how humans actually interact with the ICS and simulations to account for multiple potential behaviors.The results from both experiments show that humans react positively and more predictably when the intent of the vehicle is communicated compared to when the intent of the vehicle is unknown. In particular, the results from the simulation specifically showed a 142 percent difference between the pedestrian's trust in the vehicle's actions when the ICS is enabled and the pedestrian has prior knowledge of the vehicle than when the ICS is not enabled and the pedestrian having no prior knowledge of the vehicle.

\end{abstract}

\IEEEpeerreviewmaketitle
\section{Introduction}
Autonomous vehicles need to interact with pedestrians whenever they encounter them at crosswalks, intersections, or anytime they wander into the vehicle's path. When a human driver has this type of encounter with pedestrians, they usually provide some kind of signal, such as waving their hand, looking in to the pedestrian's eyes, or simply a smile, to let the pedestrian know they have been acknowledged. This bi-modal communication is a critical component that autonomous vehicles lack. How do they get the same point across to pedestrians without getting into a deadlock situation where neither the pedestrian nor the vehicle moves?

\begin{figure}[tbh]
	\centering
	\includegraphics[width=\columnwidth]{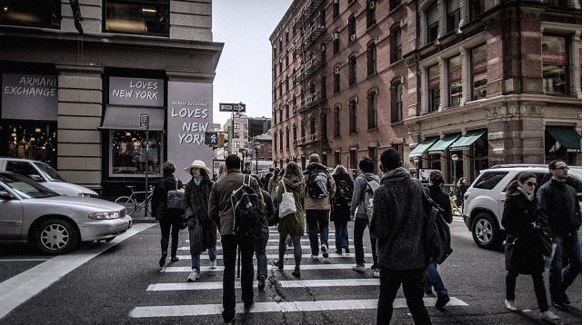}
	\caption{\label{fig:my-label} Example of how pedestrians interact with cars}
\end{figure}

\par
One of the biggest questions within the vehicle-pedestrian interaction problem is how to successfully communicate the intent of the vehicle with the surrounding pedestrians in a way that is efficient, comfortable, and easy to understand. In this case, the ability to communicate intent means that the vehicle is able to make a decision about what the pedestrian will do and then send a message to the pedestrian to try and guide their behavior as to avoid a deadlock situation. There are many researchers currently working on autonomous vehicles, along with major robotics companies, such as  Google, Tesla, and other auto manufacturers. Yet, vehicle to pedestrian communication is still an area that is being developed. Some examples of ongoing research will be discussed in the next section. However, the authors are not aware of any quantitative experiments that analyze the utility of an Intent Communication System (ICS) with real vehicle-pedestrian interactions. This paper seeks to address this gap in the literature. 

\par
The hypothesis is that if the vehicle communicates its intent to the pedestrian, this limits the number of potential actions a pedestrian may take thereby removing a large amount of uncertainty from the problem. This paper seeks to address the vehicle-pedestrian intent communication issue by examining psychological aspects of this type of communication and designing a physical and mathematical system around them. As such, our main contribution is in the design of software and hardware for an ICS and its evaluation in realistic pedestrian-vehicle encounters. The developed systems are assessed through two types of experiments: real world testing and simulations. 

\par
To conduct the real world experiments, a golf car has been outfitted with sensors that give it autonomous capabilities as well as the ability to communicate with pedestrians through an ICS. Our results suggest that this system is the first real-world evaluated and publicly reported system to help form trust between the vehicle and the humans that have to interact with it. The development of this system took careful design based on user-interaction surveys to take into consideration the environment it would operate in. It had to be simple and yet effective, reliable in dusty, water, and debris prone road situations, and not overly costly. The development of the system will be discussed further in Section 3. The simulations are an extension to the real world testing in that they allow for more scenarios to be performed that would be potentially unsafe for our participants. The simulation development will be discussed in detail in Section 4.

\par
Along with testing the ICS for communication efficiency, the paper also explores some of the psychology that is behind the perception of autonomous vehicles, what people's expectations are, and how they believe that autonomy is being integrated into society. The psychological underlyings are just as important as the physical and mathematical systems because understanding people's preferences and aversions can guide the development of more dependable and socially acceptable autonomous vehicles and in other robots that work with humans. 

\par
Summarizing, the key contributions of this paper are: a robotic intent communication system to diffuse a deadlock situation between an autonomous vehicle and pedestrian, which could prevent the pedestrian and vehicle from getting to their desired locations and a mathematical model of how the vehicle's decision making affects the pedestrian's trust. We provide details of the ICS, develop a mathematical model that shows how trust can be quantitatively defined in context of the ICS, and report detailed evaluation through experiments. 

\section{Related Work}
\par
The intent communication problem has been gaining more attention in the past year with Google, Tesla, and other automotive manufactures dedicating more resources to their autonomous vehicle development. Last year, we began development of our ICS and published a paper \cite{matthewsintent} months before Google made public a patent for their ICS \cite{urmson2015pedestrian}. Google's interest in this patent shows that the pedestrian intent communication problem is one that will need to be further researched and addressed to be able to handle the changing expectations people have of how interactions with autonomous vehicles should be. Yet, technology and fancy computer displays are not the bane of this problem; it is the careful evaluation using human feedback that is most important in designing usable ICS.  As more effort is focused on this area, more of the psychological aspects of pedestrians will have to be taken into account such as, how people feel about autonomous vehicles in general, how open people are to listen to these vehicles, and what can be done to ensure that a relationship of trust and safety can be built between humans and autonomous vehicles. Some of the issues with Google's idea is that they haven't revealed if they've done testing to determine the best way to communicate with people, how robust their system is, or how cost effective their solution is. 

\par
There are numerous researchers who are currently learning more about the psychology behind human-machine interactions (HMI). Such as in \cite{shah2010empirical}, \cite{hoc2009cooperation}, \cite{fong2006preliminary}, \cite{ivaldi2012perception}, \cite{szafir2014communication}, \cite{shah2011improved}, \cite{dragan2013legibility} which addresses the issues of how humans and machines interact with each other when compared to how humans interact with other humans. By building off of human-human interactions, autonomous systems in industrial areas like manufacturing plants or space applications are evolving to the point where they can be relied on as teammates instead of replacements. With this approach to integrating autonomy, humans have been shown to be more receptive to robotic instructions when there is some dialogue taking place between human and machine \cite{bickmore2001relational}. The ICS in this paper is expanding this idea to include a mathematical way of quantitatively measuring how much of a difference is seen when dialogue is present compared to when it isn't. Another factor that the previous papers haven't focused on as heavily is how well these systems are at creating trust between humans and machines. In our real world testing, we get a measurement of how trust is built between pedestrians and autonomous vehicles by conducting a post-survey that specifically asks for feedback on how trust was perceived to be affected.

\par
Most of the research that has been conducted on quantifying trust is in the e-commerce field \cite{manchala2000commerce}, \cite{duma2005dynamic}. These papers provide a foundation on how to develop a model for trust when people aren't sure if they are working with a human or a machine. Some other work that has been done on modeling trust is in variations of Markov Decision Processes (MDPs) \cite{goldman2003optimizing}, \cite{mcghan2015human}. These papers look into how variations of MDPs, such as partially observable Markov Decision Processes (POMDPs), decentralized POMDPs (Dec-POMDPs), multi-agent MDPs (MMDPs), and decentralized MDPs (Dec-MDPs), can be used to quantify trust in a way that accounts for the stochasticity in a human's potential actions \cite{becker2003transition}. This paper's work focuses on the use of the Dec-MDP framework for the simulations and will be discussed in Section 4. By having a framework to quantify trust, the development of the physical system can be driven by the findings from the simulation results to improve upon the state of the art in this research area of trust.
 
Various approaches have been studied to find the most reliable and natural way to communicate from machine to human; they include: gesture identification, audio feedback, haptic feedback (which is not directly applicable here), and other types of human-machine interfaces. Some of the most prominent work relies on gesture identification \cite{lee1996online}, \cite{croft2003estimating}, \cite{wang2001unsupervised}. The problem with gesture identification in the context of autonomous vehicles is that human gestures are not easily understood by machines and gestures require a large library to have an accuracy that makes them meaningful. There have also been other approaches to the communication problem as seen in \cite{pavlovic1997visual}, \cite{kucukyilmaz2012physical}, \cite{yi2014supporting}. With audio feedback, \cite{joosse2014sound}, \cite{clair2011speech} accounts for the way that people perceive sounds coming from a machine, but a problem with this is that the sounds have to be taken in a specific context. The difficulty that these methods have shown include the need for the human to have previous knowledge of the machine or training with the machine, the option of explicit or implicit communication, and the notion of trust. The current paper has found a method that is able to intertwine the notion of trust with both implicit and explicit communication in a way that people with no previous interactions with the machine will understand.

\section{Real-World Testing}

\subsection{Intent Communication System Description}

\begin{figure}[tbh]
	\centering
	\includegraphics[width=\columnwidth]{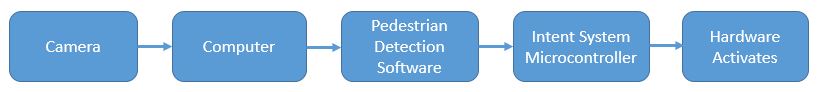}
	\caption{\label{fig:my-label} Intent Communication System Flow-Chart}
\end{figure}

\par
The intent communication system is a fusion of hardware and software. The following diagrams describe how the system is interfaced on the autonomous golf car.

\par
One of the goals with the ICS was to keep it simple, make it communicate effectively, and have the ability to operate in diverse environments. There are two strobe lights on-board the golf car, one on each side, used to get the attention of the pedestrians. The LED word display, along with the speakers, provide a message as to what the golf car would like for the pedestrian to do. Because the computers are mounted in a way where they are not seen, the cables were pointed out to show where all of the hardware is routed to. The computer contains the software used to detect pedestrians via the front-mounted camera and send a signal to the micro-controller mounted on the LED word display to tell it what message to show, how long to show it for, and when to activate the speakers and strobe lights.

\par
This system design was chosen because the goal was to ensure that the autonomous golf car was not intimidating. Studies about the Uncanny Valley have shown that the more human-like machines become, the more unsettled humans are about their appearance \cite{mori2012uncanny}.These feelings affect how people perceive technology as well as their likelihood to adopt it. There are researchers currently working on a similar system where the headlights act as eyes and they track movement. This falls under the Uncanny Valley category. Since autonomous vehicles are becoming ingrained into society through media and systems already on current vehicles such as automatic emergency braking, it is important to present the technology to them in a way that makes it easier to adapt to. 

\par
The layout chosen for this research took into account numerous factors such as, durability, cost effectiveness, the ability for the hardware to been used in varying environments, and how well this medium can relay messages. It is more robust than monitors mounted in different directions were it matters what angle the light hits them and it is not at the robot overlord end of the spectrum. 

\par
An important factor in this design also stems from the psychology behind how humans view new technology. It is essential to make people feel comfortable with new devices or they will not survive in the long term \cite{mori2012uncanny}. As people interact with these vehicles more regularly, the technology can be upgraded because they already know what to expect from the vehicle. The key is to initiate them into this new era. 

\begin{figure}[tbh]
	\centering
	\includegraphics[width=0.9\columnwidth]{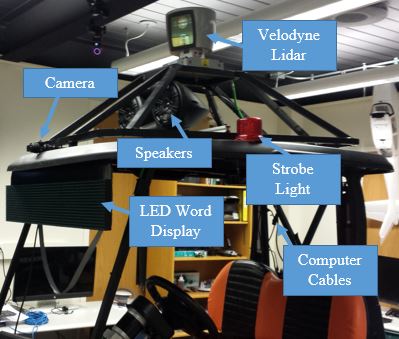}
	\caption{\label{fig:my-label} Intent Communication System components}
\end{figure}

\begin{figure}[tbh]
	\centering
	\includegraphics[width=0.9\columnwidth]{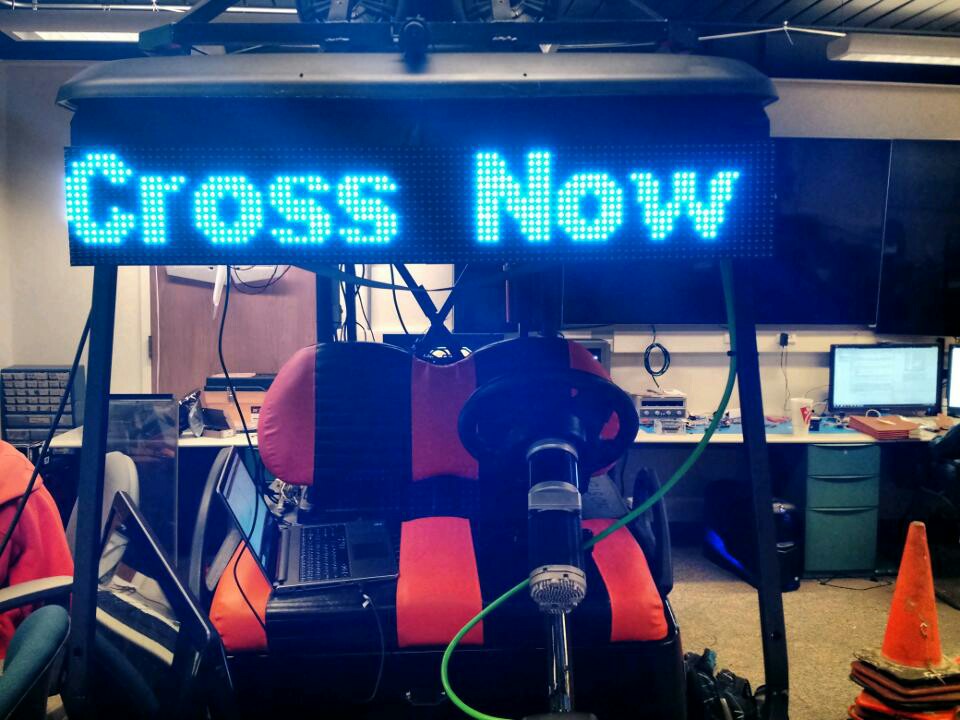}
	\caption{\label{fig:my-label} Example Visual message} 
\end{figure}

\subsection{Intent Communication Psychology}
\par
The effectiveness of the ICS has less to do with the technological sophistication of the communication system than with how people perceive the communication system. If the ICS is advanced, fancy, or complicated to the point no one understands what it is trying to do, it is useless. We found this during the initial development of the ICS when people were brought in to provide feedback on what was effective and what could be improved upon. Moving forward from this point, the biggest consideration of how to design the system was based around what humans are used to seeing on the roads. In the transition time that will follow from when autonomous cars are introduced on roads, this is a reasonable assumption. This includes things like words (including their fonts and colors), flashing lights, and sounds. 

\par
To determine what communication would be most effective, learning how humans talk to each other was crucial. In \cite{bickmore2001relational}, \cite{robinette2013building} relational trust was discussed. The dialogue that takes place between humans is an enormous source of trust building. The previously mentioned papers show how in-depth conversation is not necessary to give a command and have others follow it. For example, in an emergency situation, humans are trying to get to safety and if a robot is able to lead them to safety, they won't need in depth conversation but simple statements of what to do. Using this insight, the messages that are displayed on the ICS are as simple as saying please cross. By keeping the message short, there is little room for misinterpretation of the meaning.

\subsection{Real World Experiment}

\par
This golf car is equipped with several lidars, including a Velodyne lidar on top of the car, a front facing camera, several on-board computers, and numerous other proximity detection sensors. For these experiments, the golf car was driven by a human using a transmitter. Using the transmitter was chosen over using the autonomous capabilities due to safety precautions as the system is still under development. The human subjects were unaware of the fact that the golf car was being controlled manually, hence the results that follow are unbiased in the area of perceived autonomy. The transmitter allowed us to give the illusion of complete autonomy while still being able to drive the golf car manually and intervene if required. 

\par 
There were a total of 76 participants in the experiment: 50 in a pre-survey on actions they would take around an autonomous car and 26 subjects in the real world testing of the golf car. Between these 2 sets of participants, 9 were involved in both sets. The real world testing subjects were divided into groups as follows:

\begin{itemize}
	\item With ICS and prior knowledge of golf car (6)
	\item With ICS and no prior knowledge of golf car (7)
	\item Without ICS and prior knowledge of golf car (7)
	\item Without ICS and no prior knowledge of golf car (6)
\end{itemize}

\par
These groups were specifically chosen to reflect situations that happen in real life. There will be early adapters who want to learn as much as possible about these vehicles before everyone else and then there will be those who almost begrudgingly accept the new vehicles. In the context of these experiments, prior knowledge of the golf car means that the participants were introduced to the ICS before testing. 

\par
The participants in the prior knowledge groups were shown how everything onboard the golf car worked. They were taught how the lidars obtain data, how the ICS functioned, and more about the technology on the golf car. The participants in the no prior knowledge group didn't see the golf car until the day of testing. The groups with ICS actually had the system enabled during their interactions and those without ICS had the system disabled. The importance of changing if the ICS was enabled or not comes from how technology works in the real world. Sometimes it functions exactly how you expect and sometimes it does not and it is unclear why.

\par
None of the participants involved were aware of if there was a human inside the vehicle (there was no human in the vehicle). To explain the difference between each group, short description of each group is provided below:

\begin{itemize}
	\item With ICS and prior knowledge of golf car: this group had the ICS enabled during their encounter and they had also been introduced to the system before testing.
	\item With ICS and no prior knowledge of golf car: this group had the ICS enabled during their encounter, but they had not been introduced to the system before testing.
	\item Without ICS and prior knowledge of golf car: this group did not have the ICS enabled during their encounter, but they had been introduced to the system before testing.
	\item Without ICS and no prior knowledge of golf car: this group did not have the ICS enabled during their encounter nor had they been introduced to the system before testing.
\end{itemize}

\par
Collecting data from the experiment included a survey over how the test subjects felt before and after interacting with the golf car and video of their interactions. Some of the questions from are survey are:

\begin{itemize}
	\item Did the car do what you expected it to do?
	\item Did you behave how you expected when confronted with the car?
	\item Did the intent communication create trust in what the car would do?
	\item Did you feel safe around the car?
	\item Did you feel the communication was effective?
\end{itemize}

\par
These questions provided feedback on how the ICS was perceived and how it changed the way that the participants interacted with the golf car. While the design is important, it is critical to know how it affects perceptions of the vehicle. If there is a beautiful design and no one pays attention to what it is trying to communicate, it is an ineffective design. The survey questions were used to determine how effective the ICS was, as well as gage how it influenced the participants' trust in the vehicle's decision making abilities.

\subsection{Real World Results}

\par
During the testing event, the participants were unaware of what group they belonged to. Each participant was introduced to the experiment individually and they were given the same instructions to walk through the parking lot as if they were going to their car. They were not sure of anything the golf car would do unless they had prior knowledge of the golf car. Half of the participants had prior knowledge of the golf car so that the difference in trust could be compared to those who had never seen the golf car. Each group will be discussed separately to clearly distinguish the similarities and differences in their perceptions. 

\subsubsection{Group 1: With ICS and Prior Knowledge of Golf Car}

\par
The participants in this group were all introduced to the hardware (word display, lights, etc) on the golf car sometime prior to the experiment. When they were in the testing environment, they appeared to be the most comfortable around the golf car. Based on the answers from the survey, this group had the highest trust rating of the golf car before and after the testing. 

\par
To see how they responded to the golf car, a question about their behavior was included in the survey to gauge how much they felt their actions changed when confronted with the golf car. This group was the only group where all of the participants behaved how they expected to. The most interesting finding lies in the answer to the question: Do you feel the car is trying to replace humans or work with them? There was an approximately 50 percent split in this aspect.

\subsubsection{Group 2: With ICS and no Prior Knowledge of Golf Car}
\par
Surprisingly, this group felt that the car's actions were predictable, but they didn't trust the car more than a human driver when compared to the groups which didn't have the ICS enabled. Another contrast to Group 1 is that about 80 percent of Group 2 felt that the car was trying to work with people instead of replacing them. They had similar behaviors compared to Group 1 like, they were comfortable around the golf car and they had a relatively high trust in the golf car.

\subsubsection{Group 3: Without ICS and Prior Knowledge of Golf Car}
\par
These participants were disappointed when they didn't see the ICS during their test time. They did feel like they behaved as they expected, i.e. they thought the actions they would take were the same as the ones they did take. About 60 percent of the group felt the car was completely unpredictable and they actually felt somewhat unsafe around it. After the experiment, trust in the car was higher than before the experiment. 

\par
This was the only group where a negative change in trust was noted after exposure to the golf car without the ICS. They also believed that the ICS was made to replace humans instead of working them. The participants may also have felt like the ICS wasn't working properly. While the system was fully functional, to get a realistic range of situations that can happen in the real world, this scenario needed to be included in order to learn more about how pedestrians' trust is affected by seemingly faulty systems.

\subsubsection{Group 4: Without ICS and no Prior Knowledge of Golf Car}
\par
These participants had the lowest trust level of all the groups before and after their interactions. This result isn't surprising since they had no previous exposure to the golf car and the ICS gave them no feedback. This was the only group where most of the participants (about 80 percent) said they behaved differently than they expected to when they encountered the golf car. They didn't feel safe at all around the golf car and they felt like the purpose of the ICS was to replace human drivers. 

\begin{figure}[H]
	\begin{table}[H]
		\centering
		\caption{Real World Test Results}
		\label{my-label}
		\begin{tabular}{|l|l|l|l|l|}
			\hline
			Group Number                                    & 1  & 2  & 3  & 4  \\ \hline
			Communication was adequate                      & 28 & 24 & 19 & 9  \\ \hline
			Communication was clear                         & 28 & 26 & 9  & 7  \\ \hline
			Communication was effective                     & 29 & 25 & 13 & 8  \\ \hline
			I trust the communication of the car            & 27 & 25 & 9  & 6  \\ \hline
			I trust the car to make the appropriate actions & 25 & 23 & 15 & 12 \\ \hline
			I trust the car more because it communicates    & 28 & 22 & 8  & 6  \\ \hline
			I trust the car more than a human driver        & 18 & 15 & 19 & 9  \\ \hline
			I feel safe around the car                      & 22 & 21 & 15 & 11 \\ \hline
		\end{tabular}
	\end{table}
	\caption{Group 1: With intent communication system and prior knowledge of golf car, Group 2: With intent communication system and no prior knowledge of golf car, Group 3: Without intent communication system and prior knowledge of golf car, Group 4: Without intent communication system and no prior knowledge of golf car}
\end{figure}

\par
The results in Table 1 are based on the information gathered from the surveys after the testing was complete. To account for some groups having more participants than others, the score was scaled so the maximum number of points a group could give would be 30 because this is the maximum number of points that could be generated by the smaller groups. As seen in the table, Groups 1 and 2 had higher trust values overall compared to Groups 3 and 4. 

\par
The testing took place in a parking lot in the evening, so the participants weren't able to see inside of the golf car clearly to notice if their was someone controlling it. By keeping all of the participants blind to which group they were in, it was easier to keep the experiment unbiased from an exposure perspective. Another method used to study how people reacted to the golf car was filming the test. After reviewing the video, it was seen that the Groups 3 and 4 were more hesitant to walk in front of the golf car. They typically moved a little further away from the golf car than Groups 1 and 2. 

\par
Groups 3 and 4 were also more likely to look at the people around them for a while before they made their first step. They also were more likely to walk faster than the Groups 1 and 2. On average, Group 1 had an interaction time of 0.24 seconds before they crossed in front of the golf car during testing. Group 2 had an average interaction time of 0.61 seconds, Group 3 had an average of 0.39 seconds, and Group 4 had an average of 1.01 seconds. Interaction means the process it takes from the initial pedestrian-vehicle encounter including, the pedestrians looking at the car, determining what their best action was according to the information relayed to them by the car, and crossing in front of the car, to the final action where the pedestrian and vehicle go different ways.

\par
Groups 1 and 2 were more likely to get closer to the golf car and they spent more time observing the golf car as they walked by it, unlike Groups 3 and 4 where they would look at the golf car very briefly  as they walked by it during the test. The participants in Groups 1 and 2 appeared more confident when they crossed in front of the golf car. Some were a little startled when the display turned on, but they only paused long enough to read it and they kept moving. They seldom turned around to see what the other participants were doing. Those in Group 1 were almost over-confident as they sometimes didn't even take the time to look at the golf car as they crossed in front of it. 

\section{Simulation Testing}

\subsection{Simulation Setup}
\par
The simulations were used as a way to further study how people interact with an autonomous vehicle without the concern for safety of real participants. The simulations were run based on the same scenarios as the real world experiment. In the simulations as well as the real world experiment, there was only one pedestrian introduced to the car at a time. Only one pedestrian was chosen because pedestrians usually move in a group or there is just one pedestrian interacting with the vehicle. Groups of pedestrians are modeled as one pedestrian because these groups typically exhibit group behavior \cite{papadimitriou2016towards}, \cite{de2009pedestrian}. 

\par
Usually when one group member decides to cross and does not receive any harm, the rest of the group follows and the vehicle has to wait until it is clear to act, just as it would with one pedestrian \cite{papadimitriou2016introducing}. The simulations were based on the Dec-MDP model described later in this section. The action set of the car included: forward, stop, left, right and the action set of the pedestrian included: forward, backwards, left, right, wait, get in car, don't notice car, stop. The difference between stop and wait is that the wait action means the pedestrian might never move if the car never moves whereas with stop, the pedestrian is only waiting temporarily. 

\par
The pedestrian's action set was derived from the pre-surveys where people gave their opinions on the action they would potentially take. The survey method was chosen over just driving the car around because manufactures will have to plan for many possible pedestrian actions. Some of these actions might not arise in normal interactions and leave the vehicle and pedestrian in a deadlock situation. When people are asked to speculate as to what they might do, the answers give a broad range of actions \cite{kaparias2012analysing}.

\par
The connection between the simulations and real world experiments is that the pre-survey results were used to determine what psychological factors went into the decisions that pedestrians used to determine what action would be appropriate. 

\par
The focal point of the simulation is when there is an interaction between the car and the pedestrian. The other components of the simulation, such as when the car and pedestrian are moving about independently, are factored in where trust is considered. Trust consideration is the link between the real world experiments, simulations, and the mathematical formulation. In the context of this paper, trust is defined as the amount of confidence that the pedestrian has in the vehicle. The results from the post-survey were used to update the pedestrian model in the simulations. 

\subsection{Simulation Methodology}
\par
The goal of this research was to be able to effectively communicate with the pedestrian based on observed actions, in both the real world experiment and the simulations, and develop a mathematical model for the quantification of trust in this situation. To do this successfully, a model had to be created that incorporated the stochasticity of a pedestrian's actions, a way to measure how trust changes over time between the autonomous car and the pedestrian, and a method for being able to show how effective the communication of the car to the pedestrian was. 

\subsection{Pedestrian Surveys}

\par
To construct a realistic pedestrian model, a survey of human participants was conducted to determine the potential actions pedestrians might take when faced with an autonomous golf car and the probabilities of those actions. The survey method was chosen over observing people's reactions by driving the autonomous vehicle around due to safety precautions. Details of the survey are presented below.

\par
Based on the survey of 50 people with no affiliation to the research, potential pedestrian actions were gathered and their probabilities were calculated. The goal of the survey was to narrow down the number of actions pedestrians might take when they encounter a golf car that may or may not have a person operating it, and to use their responses to determine what kind of communication system would perform the task well and help people to feel more receptive of autonomous vehicles, so that they will be more likely to perform what the vehicle asks of them. Some of the questions asked include:

\begin{itemize}
	\item What would you do if you saw a car approaching you without a driver?
	\item How would this make you feel?
	\item What would make you feel more comfortable around this type of vehicle?
\end{itemize}

\par
The demographic of the survey included students between the ages of 18-22, faculty between the ages of 30-70, and people with no affiliation to the university between the ages of 14-65. There were five pedestrian actions identified: move, stop, wait, get in the car, and didn't notice the car. Using these actions, the Dec-MDP model was built \cite{becker2004decentralized}.

\par
The participants were specifically asked what action would they take if they saw a driverless car approaching them. They were not presented with a list of possibilities. They came up with their action based solely off of this question and an explanation of what a driverless car is. 

\par
The description of the driverless car was: a car that is able to maneuver on streets just like a car with a driver controlling it, except it relies on sensors, programs that control the decision-making capabilities, and other hardware. Some people asked for a simplified definition and they were told that the driverless car was just a car that was able to drive around without someone controlling it. Once they understood what a driverless car was, then they were able to come up with some ideas of what they would do. 

\par
The actions found were all variations of the following: stop, wait, cross, get in the car, or don't notice the car. These actions were derived from responses such as: "I don't think I'd notice it just like with a regular car", "I'd try to get in it and stop it from hitting people", and "I think I would just try to stay out of the way and wait for it to go". 

\subsection{Dec-MDP Definition}

\par
The intent of pedestrians has uncertainty associated with it due to the unpredictability of human actions. Several models were compared for their applicability to this issue. The partially observable Markov Decision Process (POMDP) was the initial choice based on the fact that the autonomous vehicle-pedestrian interaction is unknown. This means that both the mental states of the two agents is unknown and can only be found through observations. In \cite{karami2010human}, \cite{matignon2010model}, the authors discuss the limitations of using a POMDP model due to its intractability. The decentralized POMDP can be NEXP-Complete, hence suffering from computational limitations onboard the vehicle. 

\par
On the other hand, Markov Decision Processes (MDP) themselves are P-complete, however the main objection to using MDPs has been that the MDP model would not be able to account for the fact that both the car and pedestrian were unaware of the other's current and future states. In particular, the internal \textit{mental} state of both the pedestrian and the vehicle are not fully known to both the agents.  Note here that the observation of the current \textit{physical} state of both the agents is on the other hand a reasonable assumption, due to the fact that in a real environment, a car and pedestrian would be able to sense each other and see where they are with respect to one another. 

\par
The model had to be updated to include observability of at least the current state for intent communication to be effective. To overcome the issue of unobservable intent, we build on the hypothesis that if the vehicle can communicate its intent to the pedestrian, the pedestrian's actions can be reasonably well predicted. This is a key insight to making the solution implementable on real-world robots, because it means that if the ICS is active, we can effectively treat the partially observable problem as a Dec-MDP. 

\par
The Dec-MDP framework allows the notion of joint full observability. This means that the pair of observations made by both agents together fully determines the current state.  In other words, we use the ICS as a method to make it easier to communicate the current internal state of the vehicle, which is already known to the Dec-MDP, to the pedestrian and as a result, narrow the number of potential actions of the pedestrian. The situation is rather akin to what would happen if a human driver motioned for a person to cross the street. In that case, the pedestrian being aware of the driver's mind state is highly likely to cross the street.

\par
This allows us to use the Dec-MDP as a model for the vehicle-pedestrian encounter by allowing each agent to move freely, only consider each other when they intersect, and leverage the physical-state observability, like in the real world. The Dec-MDP model was appropriate for this situation because in reality, we have complete control over the car, but not the pedestrian and the model easily allows for stochasticity in one agent, while reducing the complexity by leveraging the deterministic predictability in the other, and allowing both to interact in stochastic ways that can be quantitatively measured, which is the purpose of performing the simulations. 

\par
The simulations allow more testing on many different situations involving action sets that can be changed to observe how pedestrians might behave, including unsafe situations where the pedestrian could be impaired and behave abnormally. The most important part of the simulations is that it allows us to track how trust changes over time in the duration of the interactions. This quantitative measure can be used to update the ICS to accommodate pedestrian feelings of mistrust or misunderstanding.

\par
In \cite{mcghan2012human}, \cite{mcghan2015human}, \cite{becker2004decentralized}, \cite{jennings1995controlling}, the decentralized MDP (Dec-MDP) is used in a way the accounts for stochastic behavior in an agent while showing that the need for observability can be handled. 

\begin{figure}[tbh]
	\centering
	\includegraphics[width=\columnwidth]{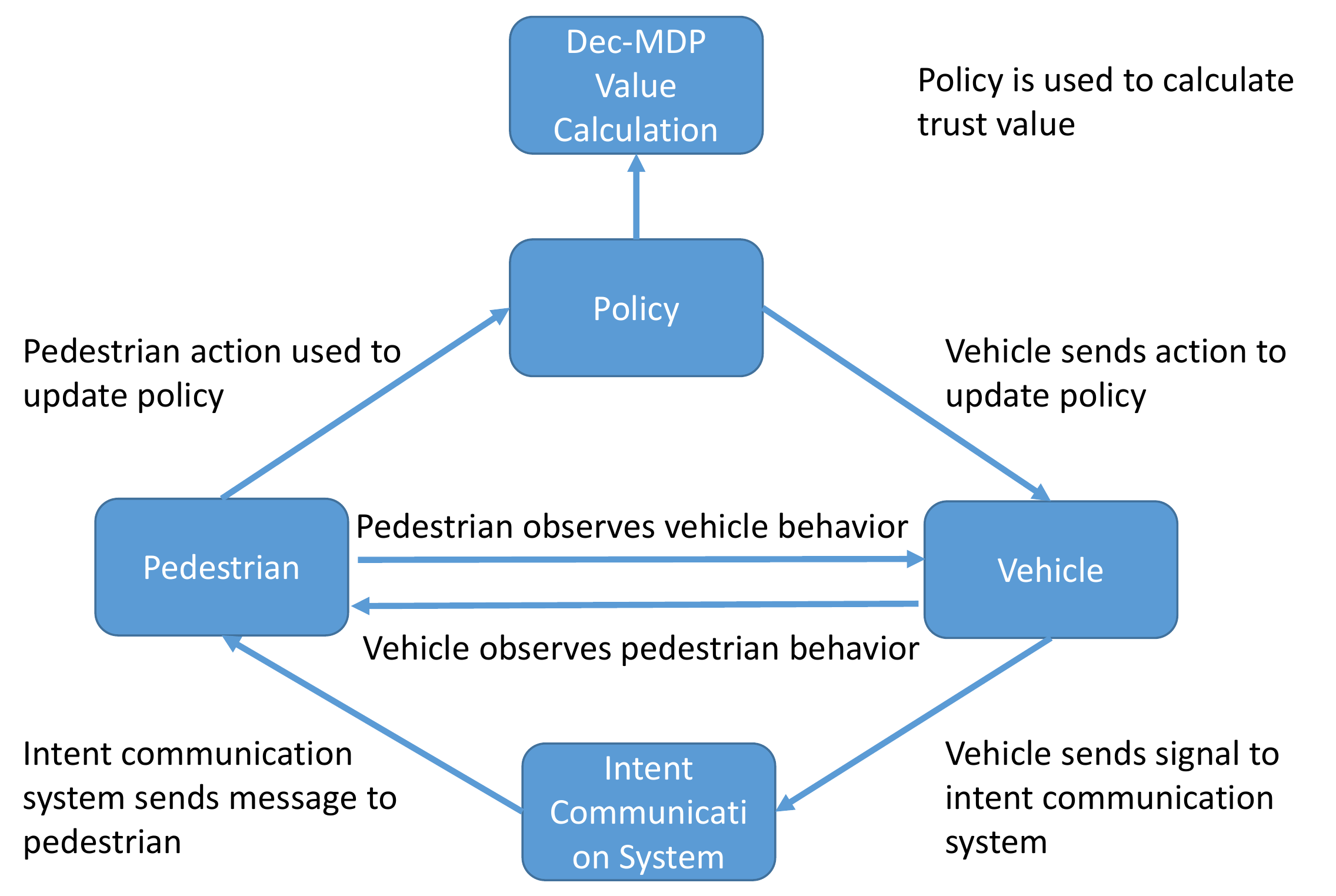}
	\caption{\label{fig:my-label} Visualization of how the Dec-MDP model works within the simulation}
\end{figure}

The Dec-MDP can be defined by the tuple $\langle S,A,P,R,O,\Omega \rangle$ where:
\begin{itemize}
	\item S is a finite set of world states both agents (car and pedestrian) share.
	
	\item A = $A_1$ $\times$ $A_2$ is a finite set of actions where $A_i$ indicates the set of actions taken by agent i.
	
	\item P is the transition function. P(s'$\mid$s, $a_1$, $a_2$) is the probability of the outcome state s' when the actions $a_1$, $a_2$ are taken in state s.
	
	\item R is the reward function. R(s, $a_1$, $a_2$, s') is the reward obtained from taking actions $a_1$, $a_2$ in state s and transitioning to state s'.
	
	\item O is the observation function. O(s, $a_1$, $a_2$, s', $o_1$, $o_2$) is the probability of agents 1 and 2 seeing observations $o_1$, $o_2$ respectively after the sequence s, $a_1$, $a_2$, s' occurs.
	
	\item $\Omega$ is the set of all observations for each of the agents.
	
\end{itemize}

\par
We formalize the intent communication problem as follows:
\begin{equation}
\begin{split}
\gamma \in [0, 1],\; \phi \in [0, 1] \\
r_1(s, a_1),\; r_2(s, a_2),\; d(s_1,s_2) \in [0,x]
\end{split}
\end{equation}

\par
In Eq.(1), $\gamma$ is the discount factor on the transitions that each agent will experience, $\phi$ is the discount factor on the trust quantification reward function which is a combination of both agents states and actions respectively that allows the consideration for how the two agents affect the individual trust of one another, $r_1$ is the reward that the vehicle receives depending on its state and action, and $r_2$ is the reward that the pedestrian receives depending on its state and action. 

\par
The rewards, $r_1$ and $r_2$, are independent of each other because as the agents navigate through the environment, they have to account for the decisions that they make individually before they interact with each other. The proximity function, $d(s_1, s_2)$, ranges between [0,x] based on the size of the simulation environment and the distance between agents. The value of \textit{x} can be adjusted to any size required, but for these simulations, it was set at 6 to accommodate the size of the gridworld environment.  

\par
The proximity function can be updated to accommodate larger or smaller environments or different test scenarios. The transition dependence of when the agents interact is accounted for in the trust quantification reward function. If the interactions should influence the states and actions each agent selects, the transition reward function will have either a positive or negative value to account for the interaction.

\par
Using the variables from Eq.(1), the reward functions can be built as follows:

\begin{equation}
R(s_1,s_2) = \sum_{i=1}^{n_{g_1}} \gamma_i r_1(s_1, a_1) + \sum_{i=1}^{n_{g_2}} \gamma_i r_2(s_2, a_2)
\end{equation}

\begin{equation}
R(s_1,s_2, A)=\sum_{j=1}^{n_{g_{1,2}}} \phi_j r_{1,2}(s_1, a_1, s_2, a_2)
\end{equation}

\begin{align}
& r_{1,2}(s_1, s_2, a_1, a_2)=\sum_{i=1}^{n_{g_1}} \theta_i r_{1_i}(s_1, a_1) + \sum_{j=1}^{n_{g_2}} \theta_j r_{2_j}(s_2, a_2)	\\
& \theta_1 = P({s_1}' \mid s_1, a_1)O(s, a_1, a_2, s', o_1, o_2)d(s_1, s_2)	\\
& \theta_2 = P({s_2}' \mid s_2, a_2)O(s, a_1, a_2, s', o_1, o_2)d(s_1, s_2)
\end{align}

\par
Eq.(2) corresponds to the transition reward function which is implicit to the Dec-MDP model and is included for completeness of the system description, but the more interesting equation is Eq.(3). Eq.(3) corresponds to the trust quantification reward function which describes how intent communication affects trust between an autonomous vehicle and a pedestrian. Eq.(3) will be the focus of further discussion of the problem. 

\par
Trust in these simulations reflects the numbers assigned to the actions that they have taken with respect to the car. In Eq.(3), $r_{1,2}(s_1, a_1, s_2, a_2)$ takes into account the interaction of the agents. The value of $r_{1,2}(s_1, a_1, s_2, a_2)$ comes from the results of the real world experiments seen in the previous section. The theta parameters, $\theta_1$ and $\theta_2$, are used to weight this reward value to include the affects of the individual agents' actions on the reward.  The values of $P({s_1}' \mid s_1, a_1)$ are based on the probability that the vehicle will take a certain action based on its current state. The values of $P({s_2}' \mid s_2, a_2)$ are based on the probabilities that were calculated from the pre-survey and the current state of the pedestrian. 

\par
These calculations were made by taking the number of actions that were mentioned and dividing the number of people who selected an action by the total number of people that participated in the survey. This reward is dependent on how the pedestrian agent views the vehicle agent. The reward varies based on the proximity of the agents with respect to each other. This proximity value is updated depending on which scenario is being tested. 

\par
This change in the proximity value based on the scenario is due to the different confidence levels seen in the pedestrians depending on the scenario they are in. The pedestrians that have more knowledge about the vehicle are typically more confident in the vehicle's ability to make sound decisions compared to someone who has never seen the vehicle. 

\par
To determine the action probabilities of the car, $P({s_1}' \mid s_1, a_1)$, we use the current state of both the car and pedestrian. The probability is updated based on the distance the car is away from the pedestrian which is varied by scenario. For example, if the scenario being tested is with ICS enabled and no prior knowledge, the car will have a 70 percent probability of stopping when it is within a distance factor, $d(s_1, s_2)$, of 3 unit spaces of the pedestrian. 

\par
Another example would be in the scenario without ICS enabled and prior knowledge where the car would have at least a 30 percent probability of stopping when it is within 3 unit spaces of the pedestrian. In essence, the transition probability is updated based on the distance that the car is away from the pedestrian, depending on the scenario. Typically, the shorter the distance, the higher probability there is for the car to stop. The unit of measure in real world situations would be in feet or meters, for example, but it is abstracted in the simulation to unit spaces.

\par
For clarification, prior knowledge means that the pedestrian has been introduced to the intent communication system before testing. They have seen how the exterior system activates and have some knowledge of the underlying software. 

\par
To calculate the action probabilities of the pedestrian for each scenario, $P({s_2}' \mid s_2, a_2)$, the results of the post-surveys were used. The most important factor for the transition probabilities was the scenario being tested. After the scenario was determined, then the answers were looked at for weighting. 

\par
The probability on the actions were dependent on how much confidence was seen in the survey results. For example, in the scenario with ICS enabled and prior knowledge, the results showed a high pedestrian confidence in both their behavior and the car's behavior which was discussed in the Section 3. The way confidence was interpreted numerically was from the value assigned to the questions. The yes or no questions were assigned values of one for yes and zero for no and the other questions were ranked on a scale of 1-5. 

\par
To determine what the probabilities on each action would be, the numbers from the surveys and the subtleties seen in the video were used to estimate how likely a pedestrian would be to take an action. In the with ICS enabled and prior knowledge scenario, they had a high confidence value. Considering that the confidence was high and what a rational human would do when they have high confidence, the risker actions such as get in the car or not notice the car were given higher probabilities at closer distances when compared to the scenarios that had lower confidence.

\par
To determine what parameters would be best suited for the simulations, the pre- and post-survey responses were used. Based on the pre-survey, an action set was obtained. To get the probability of a pedestrian taking a certain action, the number of responses given for a particular action was divided by the total number of responses which was 50. In order to decide how likely a pedestrian would be to take one of these actions in a given scenario, the post-surveys were consulted. 

\par
Based on the trust levels calculated from the responses, another probability distribution was calculated by assigning a higher value to the actions that already had a high probability of occurring based on the pre-survey. So basically the post-surveys were used to put a weight on the overall probability distribution. When the post-survey results were compared with one another, there was a strong pattern noticed being that the more information a person had about the vehicle, the higher their trust was in the decisions it made. By taking the scores from the 1-5 scaling category, the action probabilities were updated. 

\par
For example, in the ICS enabled and prior knowledge scenario, there is an overall trust rating of 27. Since the maximum is 30, 27 was divided by 30 (90 percent) to see how closely the pedestrians would follow the baseline probability distribution. The pre-survey created this baseline where the actions had the following probabilities: stop (23 percent), wait (22 percent), cross (25 percent), don’t notice the car (17 percent), get in the car (13 percent). 

\par
Once the trust value was known for a given scenario, the probability distribution was updated to show how the actions would be weighted. So in the case of ICS enabled and prior knowledge, the trust level is at 90 percent. Therefore the new probability distribution would assign higher values to the actions that have a lower probability because these actions are typically not taken by pedestrians who don’t know anything about the car.

\par
For example, there is a 90 percent chance the pedestrian will trust the car’s ability to make decisions. That means the actions can be weighted at a value of 0.9 more than what they previously were starting with the lower probability actions. 

\par
These actions are weighted first and their “extra” probability is taken from the higher valued actions and the higher valued probabilities share an equal probability after the lower probability actions have been adjusted. The probability distribution of actions in all of the scenarios would look like the following: With ICS enabled and prior knowledge: Get in the car ((13 x 0.9) +12) (24), don’t notice the car ((17 x 0.9) + 17) (32), wait (14), stop (14), cross (14).

\par
We also include the notion of joint full observability, meaning that the pair of observations made by both the agents (pedestrian and autonomous vehicle) together fully determine the current state. The notion of joint full observability can be extended to multiple pedestrians by using observations from interacting agents to determine the current state of each agent. If the agents aren't interacting, their current state isn't taken into account for the trust portion which is the primary focus. 

\par
For the interaction problem, several assumptions have been made. It is assumed that the car's behavior is fully known and controllable. This assumption stems from the fact that the car has been programmed to behave in the safest way possible meaning there is always full control of the car in either autonomous or manual mode. In other words, it is assumed that the vehicle is a deterministic system, and there is no stochasticity in the car. The action set of the car is limited to: forward, left, right, stop as these are the actions seen in normal driving operations. 

\par
Another assumption is that the car and the pedestrian are only interested in each other when they have an encounter, otherwise they act independently. The reward function contains the concept of trust as seen in Eq.(4), which is an expansion of Eq.(3). Both the car and the pedestrian have individual trust rewards depending on their proximity to one another and the observations they have based on the probability of the actions the other may take. Also, the Dec-MDP is considered over a finite-horizon because after the agents encounter each other, they no longer consider each other in their future actions unless they have another encounter.

\par
Algorithm 1 describes how the Dec-MDP was used in the simulations, the Dec-MDP solutions are based on \cite{becker2004decentralized}:
\begin{algorithm}[H]
	\caption{Intent Communication Algorithm} 
	\begin{algorithmic}[1]
		\Procedure{Dec-MDP}{$S,A,P,R,O,\Omega$}
		\State $A\gets A_1\times A_2$
		\State $s_1,s_2\gets S$
		\State $a_1,a_2\gets A$
		\State $R(s_i,a_i)=0, i=0, j=0$
		\Repeat
		\State $i\gets i+1, j\gets j+1$
		\For {$o_1, o_2$}
		\State Determine scenario $\in [1,4]$ 
		\State $p_1, p_2 \gets P(s'\mid s, a_1, a_2)$
		\State $a_1, a_2 \gets A$
		\State $\max_{a_1,a_2} r_{1,2}(s_1, s_2, a_1, a_2)$
		\For {$s_1,s_2$} check
		\If {$d(s_1,s_2) \leq scenario\ threshold$}
		\State Update $\theta_i, \theta_j$ using $d(s_1,s_2)$
		\EndIf
		\State $\pi [s_1,s_2] =\arg\max_{a_1,a_2} r_{1,2}$
		\EndFor
		\EndFor
		\Until{$s_1 = s_{g_1}$ or $s_2 = s_{g_2}$}
		\State \textbf{return} $\pi, R(s_i,a_i)$	
		\EndProcedure
	\end{algorithmic}
\end{algorithm}

\par
The purpose of the algorithm is to illustrate the development of an algorithm that is better than a reactive strategy. Because the Dec-MDP framework predicts over a finite interval, the vehicle can change its actions to accommodate the oncoming pedestrian before there is a chance for a collision. A reactive strategy would only inform the vehicle of a change in the environment when a pedestrian is already in range of the sensors and by then it may be too late for the vehicle to maneuver accordingly. 

\par
The goal with the Dec-MDP model is to keep the vehicle and pedestrian as safe as possible without incurring a large amount of computational overhead. The model also allows for the vehicle to update its actions when the pedestrian acts in an unexpected manner by giving it the observation ability. This is significantly more robust than a rule-based technique because there are a finite number of pedestrian actions that can be thought of in advance. The proximity measure $d(s_1,s_2)$ also gives more flexibility in the area of safety and trust as there is an established and adaptable distance away from the pedestrian that the vehicle starts changing its actions to accommodate pedestrian behavior.

\subsection{Simulation Results}
\par
The simulation results from the Dec-MDP model followed what was seen in the real world experiments by modeling the human trust factor after the results taken from the post-surveys. The simulation was based on a gridworld area in which only one vehicle and one pedestrian were placed. A detailed explanation of each scenario follows below. To accurately represent each scenario, parameters for the pedestrian in the simulation, such as, distance away from vehicle's effect on trust, probabilities on pedestrian actions, and the pedestrian's likelihood of following the vehicle's directions. These parameters were changed according to the scenario being tested. 

\par
The probabilities on the pedestrian's actions and the likelihood of the pedestrian following the vehicle's directions were changed to reflect what was seen in the real world experiment and the trust is modeled after the results taken from the post-surveys of the participants from the real world experiments to get as close as possible to real human variations in trust. None of the vehicle's parameters were changed due to the fact that it's actions or probabilities are the same in all scenarios. 

\par
The human model used in the simulation of each scenario varies depending on the probability distribution that was calculated based on the calculations described previously in an example. In the simulations, the pedestrian moves through a gridworld environment. The pedestrian's actual movements are like a particle in the environment. We the pedestrian decides to go to the left, it does not turn and face left, it immediately moves to the left.

\subsubsection{With ICS and Prior Knowledge}

\par	
To appropriately model this scenario, the distance away from the vehicle was reduced to one unit space. The results, seen in the plot below, show how trust was affected throughout the simulation run.

\par
As the pedestrian and vehicle interacted, the value of the trust fluctuated accordingly. As the two came in closer proximity to each other, the trust was lower than when they were further apart. The pedestrian in this scenario took more risky actions than the pedestrians in the other scenarios. In one instance, the pedestrian actually got inside of the vehicle.

\subsubsection{With ICS and No Prior Knowledge}

\par
In this scenario, the distance away from the vehicle was updated to be three unit spaces to include a higher factor of safety to account for the lack of previous knowledge.

\par
A striking difference between this scenario and the scenario where there is prior knowledge is the drastic reduction in trust. As soon as the vehicle starts moving, the pedestrian loses trust. The pedestrian here also takes some bold actions and sometimes doesn't notice the vehicle.

\subsubsection{Without ICS and Prior Knowledge}
\par
This scenario was the most interesting for updating the parameters. Like in the real world experiment, the pedestrian will be expecting commands from the vehicle, therefore the distance away from the vehicle will be the same as in the first scenario. The distance was set for one unit space. 

\par
The results from this scenario were the most unexpected. The trust value was never consistent. During some encounters between the pedestrian and vehicle, the pedestrian would have a higher trust value than at other encounters.

\subsubsection{Without ICS and Prior Knowledge}
\par
Since this scenario is the one where the pedestrian has no prior knowledge, the distance away from the vehicle is taken to the maximum of six unit spaces.

\par
Like its real world experiment counterpart, trust here is always low. It seldom moves in the positive direction and when it does, the gains are almost negligible compared to the overall trust value. All of the simulation results stop at 10 iterations because this is amount of time it takes either the vehicle or pedestrian to reach their goal. Each scenario had 2000 runs before results were analyzed. The following results are the average of the trust values for all of the runs in each scenario.

\begin{figure}[H]
	\centering
	\includegraphics[width=\columnwidth]{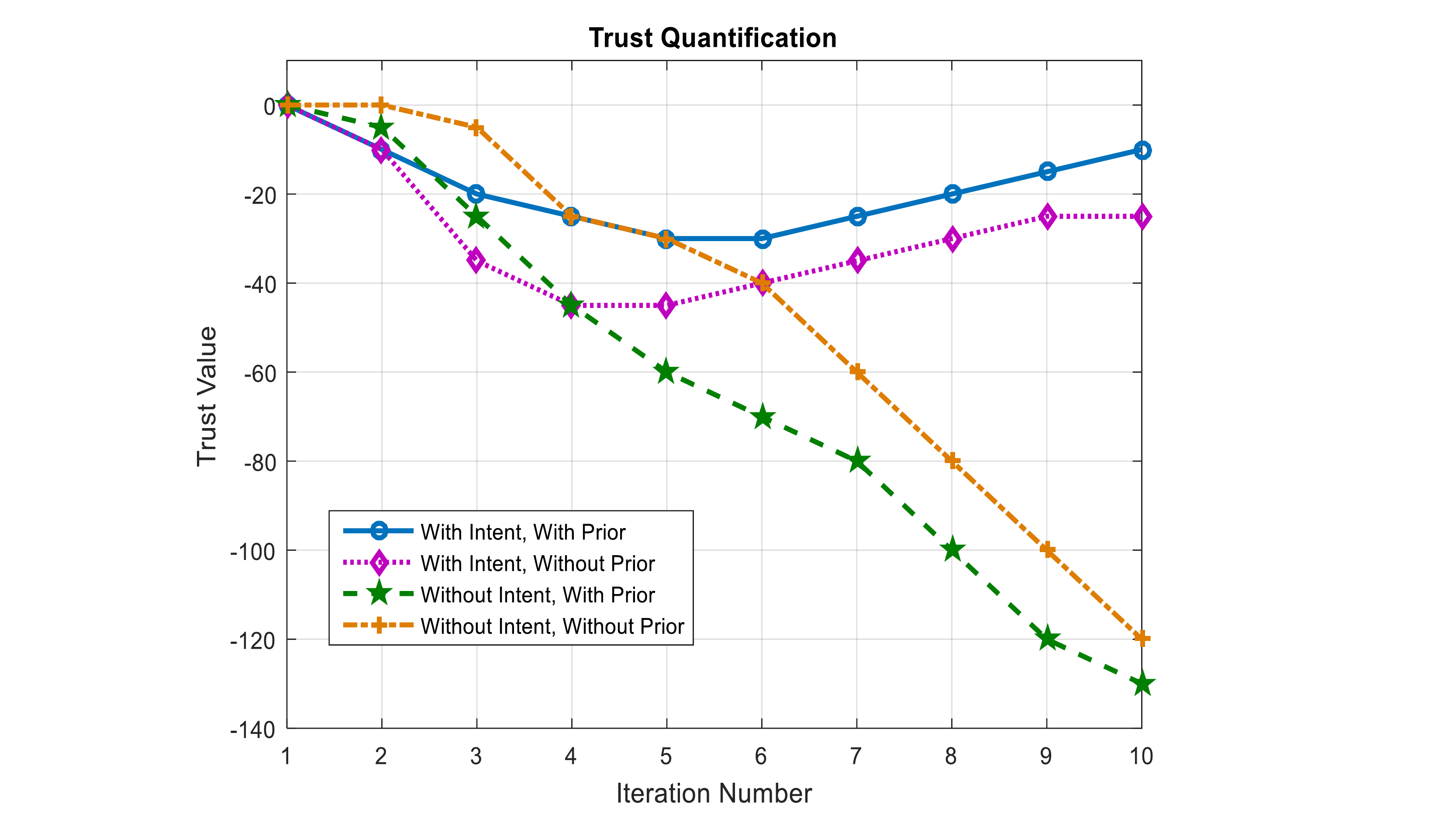}
	\caption{\label{fig:my-label} Trust quantification plot comparing all four scenarios discussed above.}
\end{figure}

\begin{figure}[H]
	\centering
	\includegraphics[width=\columnwidth]{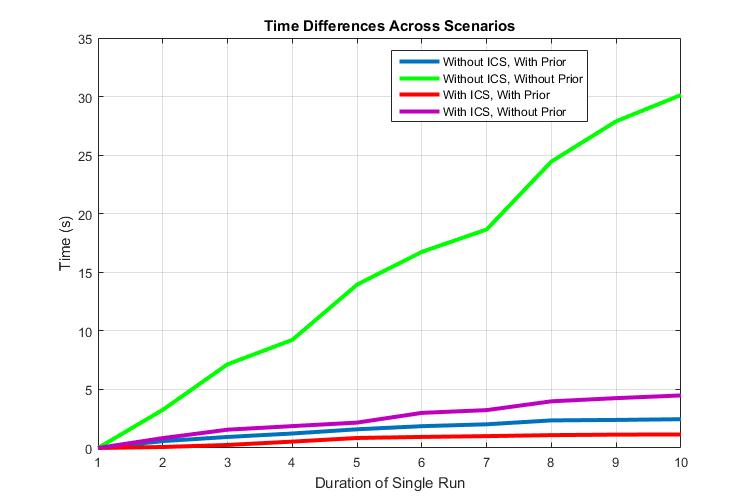}
	\caption{\label{fig:my-label} Time plot comparing all four scenarios discussed above.}
\end{figure}

\par
Because there wasn't a concrete way to assess different confidence levels in the real world testing, multiple confidence levels were considered in the simulations. The confidence term means the amount confidence the pedestrian has in the vehicle's decision making accuracy. To determine how confidence level affects the trust values in the simulations in the different scenarios, the upper and lower bounds on the results were calculated. The equation used to calculate the bounds on the results was:

\begin{equation}
C = b \pm t \times \sqrt{S}
\end{equation}

where \textit{b} is the coefficient produced by the curve fit to the data, \textit{t} is the confidence level, and \textit{S} is the is a vector of the diagonal elements from the estimated covariance matrix of the coefficient estimates, $(X^TX)^{-1}s^2$. \textit{X} is the design matrix and \textit{$s^2$} is the mean squared error. The design matrix will follow the simple regression model because there is only one explanatory variable, trust value, with several observations in each scenario. The design matrix is a matrix of two columns, the first column being ones to allow for the estimation of the y-intercept while the second column contains the x-values associated with the corresponding y-values. For each scenario, the design matrix is based on the x-values from the simulations. 

There were five confidence levels tested for each of the four scenarios: 20 percent, 40 percent, 60 percent, 80 percent, and 99 percent. These confidence levels reflect varying degrees of the pedestrian's trust in the vehicle's actions. The confidence levels can also be described as a weighting on the trust values. The following plots group together each scenario at the stated confidence level.

\begin{figure}[tbh]
	\centering
	\includegraphics[width=\columnwidth]{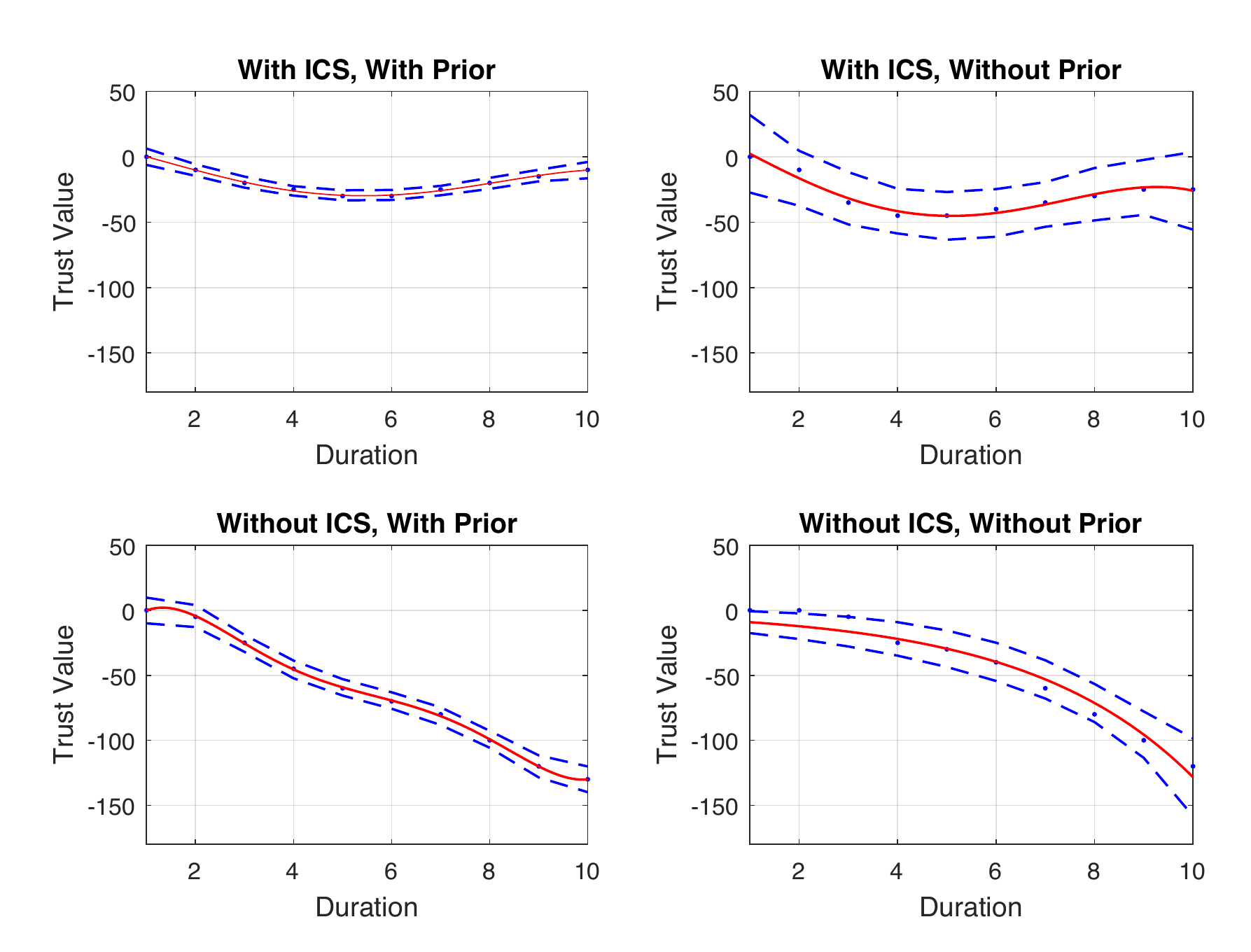}
	\caption{\label{fig:my-label} 20 Percent Level of Confidence. This is where the pedestrian has a 20 percent belief that the vehicle will make the appropriate decision.}
\end{figure}

\begin{figure}[tbh]
	\centering
	\includegraphics[width=\columnwidth]{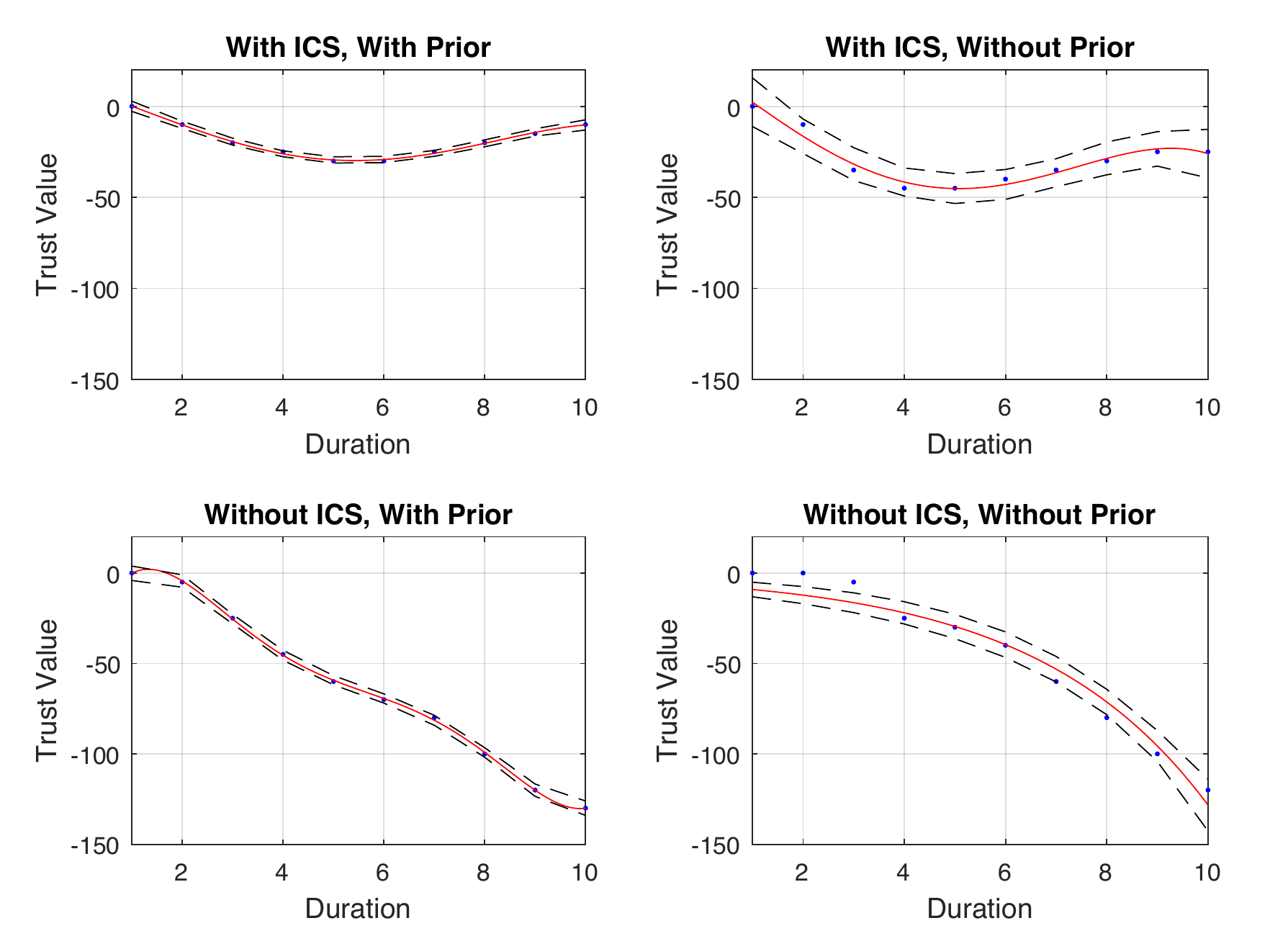}
	\caption{\label{fig:my-label} 40 Percent Level of Confidence. This is where the pedestrian has a 40 percent belief that the vehicle will make the appropriate decision.}
\end{figure}

\begin{figure}[tbh]
	\centering
	\includegraphics[width=\columnwidth]{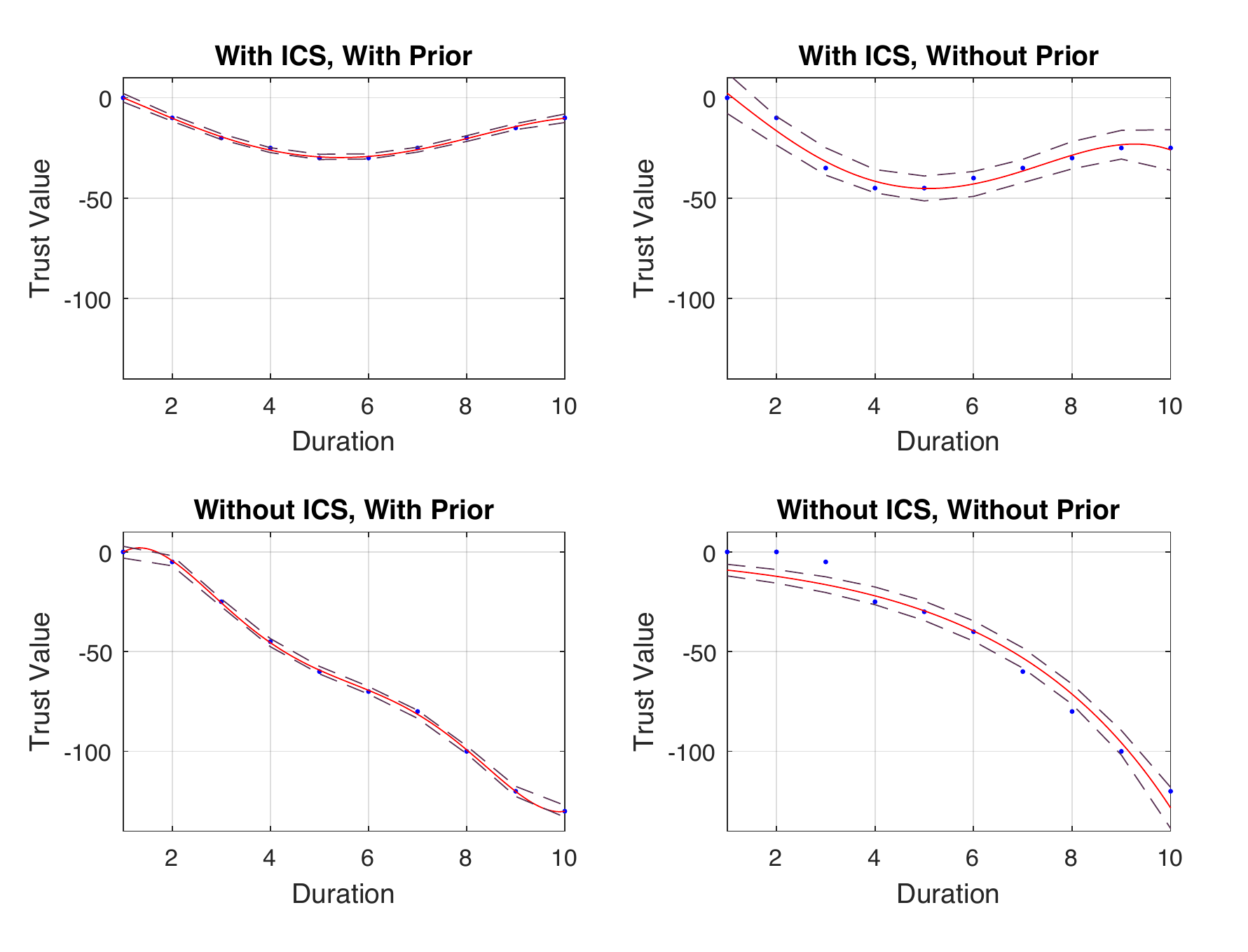}
	\caption{\label{fig:my-label} 60 Percent Level of Confidence. This is where the pedestrian has a 60 percent belief that the vehicle will make the appropriate decision.}
\end{figure}

\begin{figure}[tbh]
	\centering
	\includegraphics[width=\columnwidth]{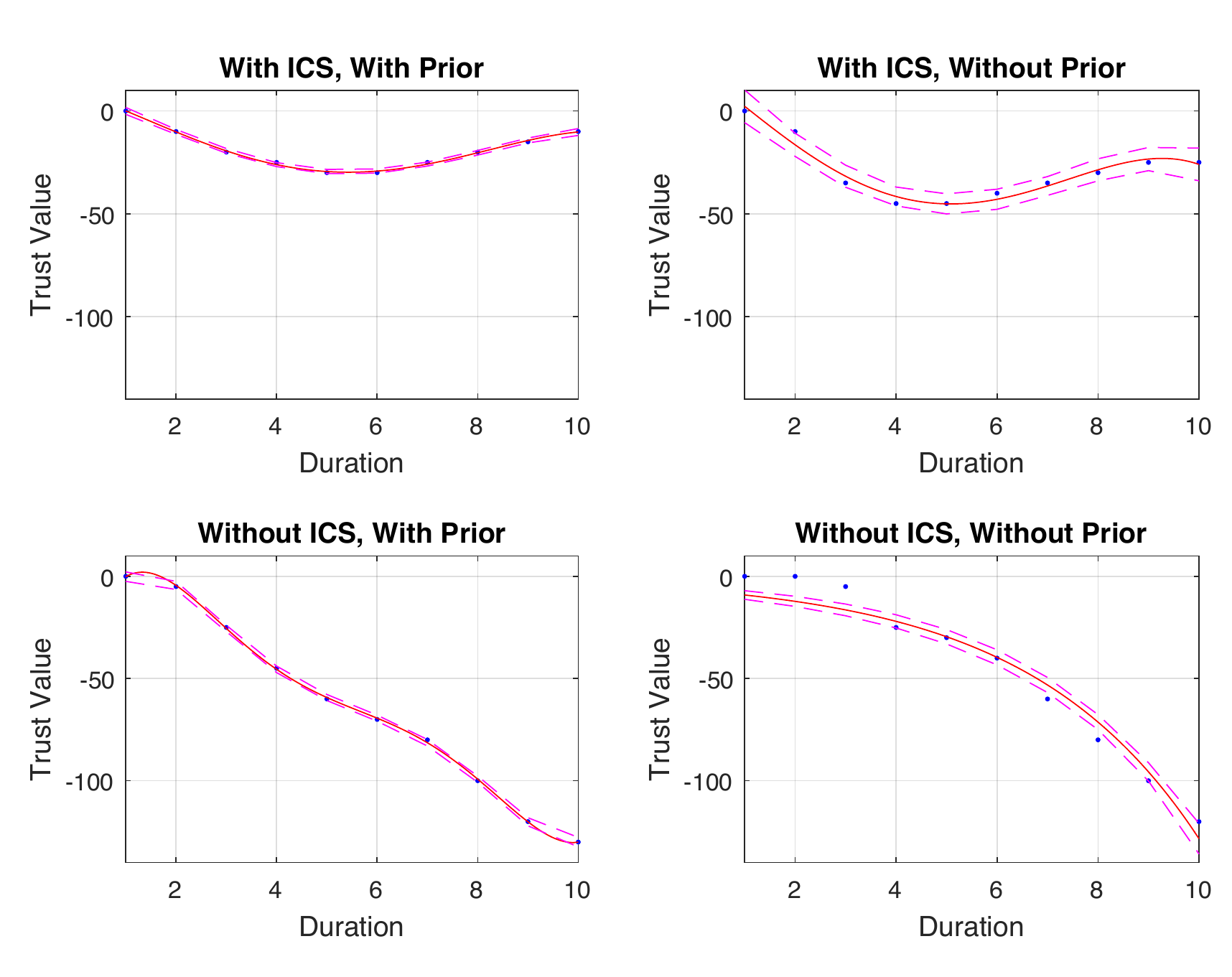}
	\caption{\label{fig:my-label} 80 Percent Level of Confidence. This is where the pedestrian has a 80 percent belief that the vehicle will make the appropriate decision.}
\end{figure}

\begin{figure}[tbh]
	\centering
	\includegraphics[width=\columnwidth]{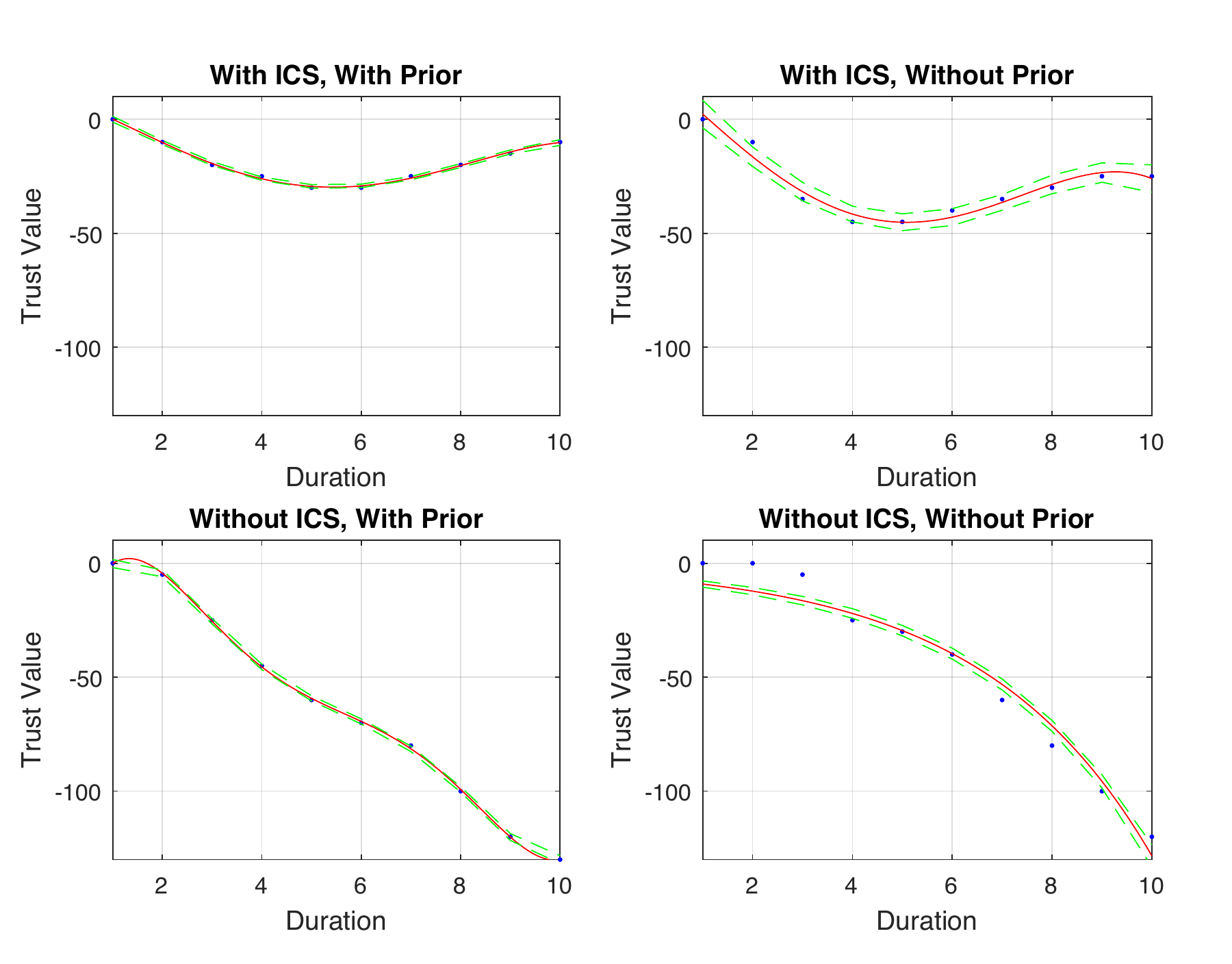}
	\caption{\label{fig:my-label} 99 Percent Level of Confidence. This is where the pedestrian has a 99 percent belief that the vehicle will make the appropriate decision.}
\end{figure}

\par
The confidence level plots reveal that as the pedestrian's confidence improves, the trust values have less variance from the baseline. The largest variance is seen in the With ICS, Without Prior scenario, while the smallest variance is seen in the Without ICS, With Prior scenario.

\par
The most interesting finding from the simulations was that no matter what scenario the pedestrian was in, the trust value was always negative with some having larger trust gains than others. This reinforces what was seen in the real world experiments. Although the range of actions was larger in the simulation, the real world participants showed a form of hesitation at some point in time during their interaction with the golf car. 

\par
The survey conducted after each participant finished was a good indicator of how they viewed the golf car, but the simulations show that there may be an underlying difference between how people report they feel about autonomous cars and how they actually view them. Another interesting note is that when the time of the simulation runs of each scenario were measured, having prior knowledge of the vehicle was actually more important than having the ICS enabled. This could lead to situations where people are introduced to vehicles in order to help integrate them into everyday life.
\section{Conclusion} 

\par
Our results clearly demonstrate through experimentation that an ICS can help in resolving potentially dangerous and inefficient deadlock situations by 38 percent. Both the real world testing and simulations were designed to evaluate how trust is affected when a pedestrian encounters and autonomous vehicle. 

\par
It was seen that trust is dependent on how comfortable a human is around the vehicle, how much prior knowledge they have of the vehicle, and the distance the vehicle is away from the human, among other factors. While this is true for most machinery, this is one of the first tests involving autonomous vehicles that confirms this holds true for this type of vehicle-human interaction. In general, those individuals who had more knowledge about the workings of the intent communication system (ICS) were more likely to trust the vehicle than those who had never seen the ICS or the vehicle. 

\par
The simulations, which were based on data taken from the real world experiments, provided a safe environment to test more risky pedestrian behaviors, including the pedestrian getting into the vehicle or not noticing the vehicle. The simulations showed that as the pedestrian interacts with the vehicle, trust levels fluctuate but never leave the negative region, emphasizing the underlying skepticism people have of autonomous vehicles, since the same group of people would not have been affected in this way if they were just interacting with a regular Golf car.
 
\par
In the 4 scenarios tested, the individuals in the groups which had prior knowledge of how the ICS worked had approximately 10 percent (Group 1 compared to Group 2) and 6 percent (Group 3 compared to Group 4) higher trust in the golf car than the groups who hadn't seen the ICS before the experiment. 

\par
The groups who had the ICS enabled, regardless of prior exposure had a 33 percent (Group 1 compared to Group 3) and 24 percent (Group 2 compared to Group 4) higher trust in the golf car than the groups that didn't have the ICS enabled. It was also seen in both real-world experiments and simulations that the groups that had prior knowledge of the system or had the ICS enabled had shorter interaction times than the groups that did not have prior knowledge or the ICS disabled.
\par
The study after the real world experiment provided crucial information about how people perceive autonomous vehicles. The groups who had the ICS enabled were more likely to trust the vehicle, but they also had slight hesitation in crossing in front of it. The groups that had the ICS disabled were more likely to not trust the vehicle and they had considerably higher hesitation. Some of the key findings were that when people were introduced to the technology beforehand, they were more likely to trust it, people had a different view on the vehicle before and after interacting with it, and they sometimes felt like the ICS was a threat because they saw it as a replacement for human drivers.


Areas where improvement can be made include: a better study to understand how to increase trust between pedestrians and autonomous vehicles, introduce more than one pedestrian at a time into experiments in either real world tests or simulations, and to study what people expect from autonomous vehicles to better design systems around them.

\bibliographystyle{plain}
\bibliography{references}

\begin{thebibliography}{10}

\bibitem{becker2004decentralized}
Raphen Becker, Shlomo Zilberstein, and Victor Lesser.
\newblock Decentralized markov decision processes with event-driven
  interactions.
\newblock In {\em Proceedings of the Third International Joint Conference on
  Autonomous Agents and Multiagent Systems-Volume 1}, pages 302--309. IEEE
  Computer Society, 2004.

\bibitem{becker2003transition}
Raphen Becker, Shlomo Zilberstein, Victor Lesser, and Claudia~V Goldman.
\newblock Transition-independent decentralized markov decision processes.
\newblock In {\em Proceedings of the second international joint conference on
  Autonomous agents and multiagent systems}, pages 41--48. ACM, 2003.

\bibitem{bickmore2001relational}
Timothy Bickmore and Justine Cassell.
\newblock Relational agents: a model and implementation of building user trust.
\newblock In {\em Proceedings of the SIGCHI conference on Human factors in
  computing systems}, pages 396--403. ACM, 2001.

\bibitem{clair2011speech}
Aaron~St Clair, Amin Atrash, Ross Mead, and Maja~J Mataric.
\newblock Speech, gesture, and space: Investigating explicit and implicit
  communication in multi-human multi-robot collaborations.
\newblock In {\em AAAI Spring Symposium: Multirobot Systems and Physical Data
  Structures}, 2011.

\bibitem{croft2003estimating}
D~Kuli{\'c}~EA Croft.
\newblock Estimating intent for human-robot interaction.
\newblock In {\em IEEE International Conference on Advanced Robotics}, pages
  810--815, 2003.

\bibitem{de2009pedestrian}
Brigitte~Cambon de~Lavalette, Charles Tijus, S{\'e}bastien Poitrenaud,
  Christine Leproux, Jacques Bergeron, and Jean-Paul Thouez.
\newblock Pedestrian crossing decision-making: A situational and behavioral
  approach.
\newblock {\em Safety Science}, 47(9):1248--1253, 2009.

\bibitem{dragan2013legibility}
Anca~D Dragan, Kenton~CT Lee, and Siddhartha~S Srinivasa.
\newblock Legibility and predictability of robot motion.
\newblock In {\em Human-Robot Interaction (HRI), 2013 8th ACM/IEEE
  International Conference on}, pages 301--308. IEEE, 2013.

\bibitem{duma2005dynamic}
Claudiu Duma, Nahid Shahmehri, and Germano Caronni.
\newblock Dynamic trust metrics for peer-to-peer systems.
\newblock In {\em Database and Expert Systems Applications, 2005. Proceedings.
  Sixteenth International Workshop on}, pages 776--781. IEEE, 2005.

\bibitem{fong2006preliminary}
Terrence Fong, Jean Scholtz, Julie Shah, Lorenzo Fluckiger, Clayton Kunz, David
  Lees, John Schreiner, Michael Siegel, Laura~M Hiatt, Illah Nourbakhsh, et~al.
\newblock A preliminary study of peer-to-peer human-robot interaction.
\newblock In {\em Systems, Man and Cybernetics, 2006. SMC'06. IEEE
  International Conference on}, volume~4, pages 3198--3203. IEEE, 2006.

\bibitem{goldman2003optimizing}
Claudia~V Goldman and Shlomo Zilberstein.
\newblock Optimizing information exchange in cooperative multi-agent systems.
\newblock In {\em Proceedings of the second international joint conference on
  Autonomous agents and multiagent systems}, pages 137--144. ACM, 2003.

\bibitem{hoc2009cooperation}
Jean-Michel Hoc, Mark~S Young, and Jean-Marc Blosseville.
\newblock Cooperation between drivers and automation: implications for safety.
\newblock {\em Theoretical Issues in Ergonomics Science}, 10(2):135--160, 2009.

\bibitem{ivaldi2012perception}
Serena Ivaldi, Natalia Lyubova, Damien G{\'e}rardeaux-Viret, Alain Droniou,
  Salvatore~Maria Anzalone, Mohamed Chetouani, David Filliat, and Olivier
  Sigaud.
\newblock Perception and human interaction for developmental learning of
  objects and affordances.
\newblock In {\em Humanoid Robots (Humanoids), 2012 12th IEEE-RAS International
  Conference on}, pages 248--254. IEEE, 2012.

\bibitem{jennings1995controlling}
Nicholas~R Jennings.
\newblock Controlling cooperative problem solving in industrial multi-agent
  systems using joint intentions.
\newblock {\em Artificial intelligence}, 75(2):195--240, 1995.

\bibitem{joosse2014sound}
Michiel Joosse, Manja Lohse, and Vanessa Evers.
\newblock Sound over matter: the effects of functional noise, robot size and
  approach velocity in human-robot encounters.
\newblock In {\em Proceedings of the 2014 ACM/IEEE international conference on
  Human-robot interaction}, pages 184--185. ACM, 2014.

\bibitem{kaparias2012analysing}
Ioannis Kaparias, Michael~GH Bell, Ashkan Miri, Carol Chan, and Bill Mount.
\newblock Analysing the perceptions of pedestrians and drivers to shared space.
\newblock {\em Transportation research part F: traffic psychology and
  behaviour}, 15(3):297--310, 2012.

\bibitem{karami2010human}
A-B Karami, Laurent Jeanpierre, and A-I Mouaddib.
\newblock Human-robot collaboration for a shared mission.
\newblock In {\em Human-Robot Interaction (HRI), 2010 5th ACM/IEEE
  International Conference on}, pages 155--156. IEEE, 2010.

\bibitem{kucukyilmaz2012physical}
Ayse Kucukyilmaz, Tevfik~Metin Sezgin, and Cagatay Basdogan.
\newblock Physical communication of intent: A haptic negotiation framework for
  human-robot interaction.
\newblock 2012.

\bibitem{lee1996online}
Christopher Lee and Yangsheng Xu.
\newblock Online, interactive learning of gestures for human/robot interfaces.
\newblock In {\em Robotics and Automation, 1996. Proceedings., 1996 IEEE
  International Conference on}, volume~4, pages 2982--2987. IEEE, 1996.

\bibitem{manchala2000commerce}
Daniel~W Manchala.
\newblock E-commerce trust metrics and models.
\newblock {\em Internet Computing, IEEE}, 4(2):36--44, 2000.

\bibitem{matignon2010model}
La{\"e}titia Matignon, Abir-Beatrice Karami, and Abdel-Illah Mouaddib.
\newblock A model for verbal and non-verbal human-robot collaboration.
\newblock In {\em AAAI Fall Symposium: Dialog with Robots}, 2010.

\bibitem{matthewsintent}
Milecia Matthews and Girish~V Chowdhary.
\newblock Intent communication between autonomous vehicles and pedestrians.

\bibitem{mcghan2012human}
Catharine McGhan, Ali Nasir, and Ella Atkins.
\newblock Human intent prediction using markov decision processes.
\newblock {\em Infotech@ Aerospace}, 2012:2, 2012.

\bibitem{mcghan2015human}
Catharine~LR McGhan, Ali Nasir, and Ella~M Atkins.
\newblock Human intent prediction using markov decision processes.
\newblock {\em Journal of Aerospace Information Systems}, pages 1--5, 2015.

\bibitem{mori2012uncanny}
Masahiro Mori, Karl~F MacDorman, and Norri Kageki.
\newblock The uncanny valley [from the field].
\newblock {\em Robotics \& Automation Magazine, IEEE}, 19(2):98--100, 2012.

\bibitem{papadimitriou2016towards}
Eleonora Papadimitriou.
\newblock Towards an integrated approach of pedestrian behaviour and exposure.
\newblock {\em Accident Analysis \& Prevention}, 92:139--152, 2016.

\bibitem{papadimitriou2016introducing}
Eleonora Papadimitriou, Sylvain Lassarre, and George Yannis.
\newblock Introducing human factors in pedestrian crossing behaviour models.
\newblock {\em Transportation Research Part F: Traffic Psychology and
  Behaviour}, 36:69--82, 2016.

\bibitem{pavlovic1997visual}
Vladimir~I Pavlovic, Rajeev Sharma, and Thomas~S. Huang.
\newblock Visual interpretation of hand gestures for human-computer
  interaction: A review.
\newblock {\em Pattern Analysis and Machine Intelligence, IEEE Transactions
  on}, 19(7):677--695, 1997.

\bibitem{robinette2013building}
Paul Robinette, Alan~R Wagner, and Ayanna~M Howard.
\newblock Building and maintaining trust between humans and guidance robots in
  an emergency.
\newblock In {\em AAAI Spring Symposium: Trust and Autonomous Systems}, pages
  78--83, 2013.

\bibitem{shah2010empirical}
Julie Shah and Cynthia Breazeal.
\newblock An empirical analysis of team coordination behaviors and action
  planning with application to human--robot teaming.
\newblock {\em Human Factors: The Journal of the Human Factors and Ergonomics
  Society}, 52(2):234--245, 2010.

\bibitem{shah2011improved}
Julie Shah, James Wiken, Brian Williams, and Cynthia Breazeal.
\newblock Improved human-robot team performance using chaski, a human-inspired
  plan execution system.
\newblock In {\em Proceedings of the 6th international conference on
  Human-robot interaction}, pages 29--36. ACM, 2011.

\bibitem{szafir2014communication}
Daniel Szafir, Bilge Mutlu, and Terrence Fong.
\newblock Communication of intent in assistive free flyers.
\newblock In {\em Proceedings of the 2014 ACM/IEEE international conference on
  Human-robot interaction}, pages 358--365. ACM, 2014.

\bibitem{urmson2015pedestrian}
Christopher~Paul Urmson, Ian~James Mahon, Dmitri~A Dolgov, and Jiajun Zhu.
\newblock Pedestrian notifications, November~24 2015.
\newblock US Patent 9,196,164.

\bibitem{wang2001unsupervised}
Tian-Shu Wang, Heung-Yeung Shum, Ying-Qing Xu, and Nan-Ning Zheng.
\newblock Unsupervised analysis of human gestures.
\newblock In {\em Advances in Multimedia Information Processing—PCM 2001},
  pages 174--181. Springer, 2001.

\bibitem{yi2014supporting}
Daqing Yi and Michael~A Goodrich.
\newblock Supporting task-oriented collaboration in human-robot teams using
  semantic-based path planning.
\newblock In {\em Society of Photo-Optical Instrumentation Engineers (SPIE)
  Conference Series}, volume 9084, 2014.

\end{thebibliography}

\end{document}